\def\Title{Intersubjectivity and value reproducibility of outcomes of quantum measurements}
\def\Author{Masanao Ozawa}
\def\AffilChubu{
Center for Mathematical Science and Artificial Intelligence,
Academy of Emerging Sciences, Chubu University,
1200 Matsumoto-cho, Kasugai, 487-8501, Japan.
Email:~m.ozawa@fsc.chubu.ac.jp}
\def\AffilNagoya{
Graduate School of Informatics,
Nagoya University, Chikusa-ku, Nagoya, 464-8601, Japan.
Email:~ozawa@is.nagoya-u.ac.jp
}
\def\AffilRIKEN{
Fundamental Quantum Science Program, TRIP Headquarters, iTHEMS, RIKEN, Hirosawa, Wako, Saitama, 351-0198, Japan.
Email:~masanao.ozawa@riken.jp
}
\def\AffilRitsumei{
Kinugasa Research Organization, Ritsumeikan University, Kyoto, 603-8577, Japan.
Email:~m-ozawa@fc.ritsumei.ac.jp}
\def\Abstract{
Every measurement determines a single value as its outcome, and yet quantum mechanics predicts it only probabilistically. The Kochen-Specker theorem and Bell's inequality are often considered to reject a realist view but favor a skeptical view that measuring an observable does not mean ascertaining the value that it has, but producing the outcome, having only a personal meaning. However, precise analysis supporting this view is unknown. Here, we show that a quantum mechanical analysis turns down this view. Supposing that two observers simultaneously measure the same observable, we can well pose the question as to whether they always obtain the same outcome, or whether the probability distributions are the same, but the outcomes are uncorrelated. Contrary to the widespread view in favor of the second, we shall show that quantum mechanics predicts that only the first case occurs. We further show that any measurement establishes a time-like entanglement between the observable to be measured and the meter after the measurement, which causes the space-like entanglement between the meters of different observers. We also show that our conclusion cannot be extended to measurements of so-called `generalized' or `unsharp' observables, suggesting a demand for reconsidering the notion of observables in foundations of quantum mechanics.}
\documentclass[fleqn,10pt]{wlscirep}
\usepackage[normalem]{ulem}
\usepackage[utf8]{inputenc}
\usepackage[T1]{fontenc}
\usepackage{amsmath,amssymb,amsthm} 
\usepackage{fdsymbol}
\newtheorem{Theorem}{Theorem}
\newcommand*{\bTheorem}{\begin{Theorem}}
\newcommand*{\eTheorem}{\end{Theorem}}	
\newtheorem{Assumption}{Assumption}
\newcommand*{\bAssumption}{\begin{Assumption}}
\newcommand*{\eAssumption}{\end{Assumption}}	
\newcommand*{\benum}{\begin{enumerate}\itemsep=0in \parskip=0in}
\newcommand*{\eenum}{\end{enumerate}}
\newcommand*{\bProof}{\begin{proof}}
\newcommand*{\eProof}{\end{proof}}
\renewcommand*{\And}{\wedge}
\newcommand*{\R}{\mathbb{R}}

\newcommand*{\av}[1]{\langle#1\rangle}

\newcommand*{\da}{\dagger}
\newcommand*{\de}{\delta}

\newcommand*{\ie}{{\it i.e.}}

\newcommand*{\nn}{\nonumber}

\newcommand*{\ph}{\varphi}
\newcommand*{\ps}{\psi}

\newcommand*{\ta}{\tau}

\newcommand*{\bE}{\mathbf{E}}
\newcommand*{\bS}{\mathbf{S}}

\newcommand*{\cG}{\mathcal G}
\newcommand*{\cH}{\mathcal{H}}
\newcommand*{\cK}{\mathcal{K}}
\newcommand*{\cL}{\mathcal{L}}
\newcommand*{\cM}{\mathcal{M}}
                                          
\newcommand*{\Eq}[1]{Eq.~(\ref{eq:#1})}

\newcommand*{\Ps}{\Psi}

\newcommand*{\Then}{\Rightarrow}
\newcommand*{\bra}[1]{\langle#1|}
\newcommand*{\ket}[1]{|#1\rangle}
\newcommand*{\ketbra}[1]{\ket{#1}\bra{#1}}
\newcommand*{\kps}{\ket{\psi}}
\newcommand*{\kPs}{\ket{\Ps}}
\newcommand*{\kxi}{\ket{\xi}}
\newcommand*{\kph}{\ket{\ph}}

\newcommand*{\beq}{\begin{equation}}
\newcommand*{\eeq}{\end{equation}}         
\newcommand*{\beql}[1]{\begin{equation}\label{eq:#1}}
\newcommand*{\beqa}{\begin{eqnarray}}
\newcommand*{\eeqa}{\end{eqnarray}}
\newcommand*{\beqas}{\begin{eqnarray*}}
\newcommand*{\eeqas}{\end{eqnarray*}}
\newcommand*{\deq}[1]{ \begin{align}#1\end{align}}

\newcommand*{\deqn}[1]{ \begin{align}#1\end{align}}
\newcommand*{\deql}[2]{\begin{align}\label{eq:#1}#2\end{align}}

\title{\Title}
\author[1,2,3,4]{\Author}
\affil[1]{\AffilChubu}
\affil[2]{\AffilNagoya}
\affil[3]{\AffilRIKEN}
\affil[4]{\AffilRitsumei}
\begin{abstract}
\Abstract
\end{abstract}
\begin{document}
\flushbottom
\maketitle
\thispagestyle{empty}
\section*{Introduction -- Conventional view}
The theorems due to Kochen-Specker \cite{KS67} and Bell \cite{Bel64}, enforced by the recent loophole-free experimental tests
\cite{Hen15,Giu15,Sha15}, 
are often considered to defy the correlation between the measurement outcome and 
the {\em pre-}measurement value of the measured observable. Accordingly, it is a standard 
view that the measurement outcome should only correlate to the {\em post}-measurement value of the 
measured observable, as Schr\"{o}dinger stated long ago:
\begin{quote}
The rejection of realism has logical consequences. 
In general, a variable has no definite value before I measure it; then 
measuring it does not mean ascertaining the value that it has. But then what does it 
mean? [$\ldots$ ] 
Now it is fairly clear; if reality does not determine the measured value, 
then at least the measured value must 
determine reality [$\ldots$ ]  That is, the desired criterion can be merely this: 
repetition of the measurement must 
give the same result. \cite[p.~329]{Sch35E}
\end{quote}

The {\em repeatability hypothesis} 
mentioned above 
is formulated as one of the basic axioms
of quantum mechanics by von Neumann:
\begin{quote}
If [a] physical quantity is measured twice in succession 
in a system, then we get the same value each time. 
 \cite[p.~335]{vN32E}
\end{quote}
It is well known that this hypothesis is equivalent to the {\em collapsing hypothesis}
formulated by Dirac:
\begin{quote}
A measurement always causes the system to jump into an eigenstate of the observable
that is being measured, the eigenvalue this eigenstate belongs to being equal 
to the result of the measurement.   \cite[p.~36]{Dir58}
\end{quote}

The repeatability hypothesis and the collapsing hypothesis formulated as above had been broadly accepted since the inception of quantum 
mechanics. However, they have been abandoned in the modern formulation 
treating all the physically realizable quantum measurements.
In fact, Davies and Lewis \cite{DL70} proposed to abandon the repeatability hypothesis and introduced an ``operational approach'' leading to a more flexible approach to measurement theory based on a mathematical notion of an ``instrument''.
Subsequently, Yuen \cite{Yue87} proposed
the problem of mathematically characterizing all the possible quantum measurements claiming that the Davies--Lewis ``operational approach'' is too general.  
Interestingly, Yuen's problem had been already solved at that time by the present author \cite{84QC} showing 
that all the physically realizable quantum measurements are exactly characterized by ``completely positive instruments'', 
instruments state changes of which  satisfy complete positivity,
nowadays a broadly accepted notion often called as ``quantum instruments''.

Von Neumann \cite[p.~440]{vN32E} found a measuring interaction satisfying 
the repeatability condition for an observable 
$
A=\sum_a a\ket{\ph_a}\bra{\ph_a}.
$
He showed that such a measurement is described by a unitary operator $U(\tau)$
such that
\deql{101103a}{
U(\tau)\ket{\ph_a}\ket{\xi}=\ket{\ph_a}\ket{\xi_a},
}
where $\ket{\xi}$ is an arbitrary but fixed state of the environment,
$\{\ket{\xi_a}\}$ is an orthonormal basis for the environment,
and the meter observable is given by
$
M=\sum_a a\ketbra{\xi_a}.
$
If the initial system state is a superposition
$\kps=\sum_{a}c_a \ket{\ph_{a}}$, by linearity we obtain
\deql{U-sp}{
U(\tau)\ket{\psi}\ket{\xi}=\sum_{a}c_a\ket{\ph_{a}}\ket{\xi_{a}}.
}
Then we have
\deq{
\Pr\{A(\tau)=x,M(\tau)=y\}=\de_{x,y}|c_x|^2.
}
Thus, this measurement satisfies the {\em probability reproducibility condition}, 
\deql{PRC0}{
\Pr\{M(\tau)=x\}=|c_x|^2,
}
and the {\em repeatability condition},
\deql{RC}{
\Pr\{A(\tau)=x,M(\tau)=y\}=0\quad\mbox{if $x\ne y$};
}
see \cite[p.~440]{vN32E}.

According to the above 
analysis,
the measurement outcome is often considered
to be created, rather than reproduced,  
by the act of measurement \cite{Jor32}.  
Quantum Bayesian interpretation emphasizes its personal nature as one of the fundamental tenets \cite[Abstract]{Fuc17}:
\begin{quote}
 Along the way, we lay out three tenets of QBism in some detail: [\ldots] 
3) Quantum measurement outcomes just are personal experiences for the agent gambling upon them.
\end{quote}

However, if we consider the process of measurement from a more general perspective abandoning the repeatability
hypothesis,  in which only the probability reproducibility condition is required, we confront a puzzling problem. 

Suppose that two remote observers, I and II, simultaneously measure the same observable.  
Then, we can ask whether quantum mechanics predicts that they always obtain the same outcome, 
or quantum mechanics predicts only that their probability distributions 
are the same but the outcomes are uncorrelated.
In the following we shall show that quantum mechanics predicts that only the first case occurs, in contrast to a common interpretation 
of the theorems due to Kochen-Specker \cite{KS67} and Bell \cite{Bel64}.

\section*{Results}
\subsection*{Intersubjectivity of outcomes of quantum measurements}
It is fairly well-known that any measurement can be described 
by an interaction between the system $\bS$ to be measured and the environment 
$\bE$ including measuring apparatuses
and that the outcome of the measurement is obtained by a subsequent observation 
of a meter observable in the environment by the observer \cite{vN32E,84QC,23A2}.

Let $A$ be an observable to be measured.  
Let $M_1$ and $M_2$ be the meter observables of observers I and II, respectively.
We assume that at time 0 the system $\bS$ is in an arbitrary state  
$\kps$ and the environment $\bE$ is in a fixed state $\kxi$, respectively.
In order to measure the observable 
$A$ at time 0, observers I and II locally measure their meters $M_1$ and $M_2$ 
at times $\tau_1>0$ and $\tau_2>0$, respectively.  
  
 Then, the time evolution operator $U(t)$ of the total system $\bS+\bE$
 determines the Heisenberg operators $A(0)$,  
 $M_1(t)$, $M_2(t)$ for any time $t>0$, where
$A(0)=A\otimes I$,
$M_1(t)=U(t)^{\dagger}(I\otimes M_1) U(t)$, and  
$M_2(t)=U(t)^{\dagger}(I\otimes M_2) U(t)$.
For any observable $X$, we denote by $P^{X}(x)$ the spectral 
projection of $X$ corresponding to $x\in\R$, \ie, $P^{X}(x)$ is the projection 
onto the subspace of vectors  $\kps$ satisfying $X\kps=x\kps$.

We pose the following two assumptions.

\bAssumption[Locality]
We suppose that 
  $M_1(\tau_1)$ and $M_2(\tau_2)$ are mutually commuting and that the
   joint probability distribution 
   of the outcomes of measurements by observers I and II are 
   given by 
\deql{JPD}{
\Pr\{M_1(\tau_1)=x,M_2(\tau_2)=y\}=\av{\psi,\xi|P^{M_1(\tau_1)}(x)P^{M_2(\tau_2)}(y)|\psi,\xi}
}
for all $x,y\in\R$, where $\ket{\psi,\xi}=\ket{\psi}\ket{\xi}$.
\eAssumption

\bAssumption[Probability reproducibility] 
The measurements of the observable $A$ by observers I and II satisfy 
  the {\em probability reproducibility condition},
  \ie, 
  \deql{A2}{
  \Pr\{M_1(\ta_1)=x\}= \Pr\{M_2(\ta_2)=x\}=\Pr\{A(0)=x\}
  }
  for any $x\in\R$ in arbitrary $\kps$ and fixed $\kxi$.
\eAssumption

Assumption 1 is a natural consequence from the assumption that
the two local meter-measurements by observers I and II are space-like separated.
 In this case, the Local Measurement Theorem \cite[Theorem 5.1]{97QQ}
ensures that the joint probability distribution of 
the outcomes of measurements by observers I and II satisfies \Eq{JPD},
without assuming that the meter measurements satisfy the repeatability 
or collapsing condition.
Thus, the joint probability of their outcomes is well-defined by \Eq{JPD},
and our problem is well-posed.

In Assumption 2 we only require that the outcome of a measurement 
of an observable should satisfy the Born rule for the measured observable, 
and we make no assumption on the state change caused by the measurement
such as the repeatability condition nor the collapsing condition.

Now we can ask if observer I and II always obtain the same outcome,
\ie, 
 \deql{Main}{
 \Pr\{M_1(\tau_1)=x,M_2(\tau_2)=y\}=0
 }
if $x\ne y$.
We call this condition the {\em intersubjectivity condition}.

Now we shall show the following.
\bTheorem[Intersubjectivity Theorem] \label{th:1}
The outcomes of simultaneous, probability reproducible
measurements of the same observable with two space-like separated meter observables satisfy the intersubjectivity condition.
\eTheorem

A complete proof is given in the Methods section.
We call the condition
\deql{QPC-0}{P^{M(\ta)}(x)\kps\kxi=P^{A(0)}(x)\kps\kxi}
the {\em time-like entanglement condition}, 
comparing with the {\em space-like entanglement condition}
\deq{ 
P^{M(\ta)}(x)\kps\kxi=P^{A(\ta)}(x)\kps\kxi,
}
which follows from \Eq{U-sp}.

Note that an alternative proof for Theorem \ref{th:1} can be obtained from an advanced structure theorem \cite[Theorem 3.2.1]{Dav76} for  joint POVMs (probability operator-valued measures). In fact, if we apply this theorem to the joint POVM 
\deq{\Pi(x,y)=\av{\xi|P^{M_1(\ta_1)}(x)P^{M_2(\ta_2)}(y)|\xi},} 
we obtain the relation $\Pi(x,y)=P^{A}(x)P^{A}(y)$ and \Eq{Main} follows.
Nevertheless, the proof of Theorem \ref{th:1} in this paper clearly shows the following points, which are useful for our later discussions.
\begin{itemize}
\item The probability reproducibility condition implies the time-like entanglement condition.
\item The intersubjectivity condition is a straightforward consequence from the time-like entanglement condition for two space-like separated
meter observables.
\end{itemize}

It can be easily seen that Theorem \ref{th:1} can be extended to the assertion 
  for $n$ observers with any $n>2$.
    Thus, we conclude that {\em if two or more mutually space-like separated  observers simultaneously 
   measure  the same observable, then their outcomes always coincide. }

Example 1 in the Methods section illustrates a typical system-environment interaction 
to realize simultaneous position measurements of $n$ observers.

\section*{Non-Intersubjectivity for unconventional generalized observables}
We note that the above result, Theorem \ref{th:1},  cannot be extended to 
an arbitrary `generalized observable' $A$  represented by a POVM, \ie,  a family $\{P^{A}(x)\}_{x\in\R}$ of 
positive operators $P^{A}(x)\ge 0$, not necessarily of projections, 
such that $\sum_{x}P^{A}(x)=I$.
The optical phase is not considered as a quantum observable
but typically considered as a physical quantity corresponding 
to a generalized observable (see Ref.~\cite{97PM} 
and the references therein). 

To immediately see that our conclusion cannot be extended to the class of  
generalized observables, 
consider a generalized observable $A$ defined by $P^{A}(x)=\mu(x)I$, 
where $\mu$ is an arbitrary probability
distribution, \ie, $\mu(x)\ge 0$ and $\sum_{x}\mu(x)=1$. 
Then, as shown in Example 2 in the Methods section,
we can construct continuously parametrized models for which 
Assumptions 1 and 2 hold, but the intersubjectivity condition, \Eq{Main}, does not hold.

 \section*{Value reproducibility of outcomes of quantum measurements}
The intersubjectivity of the measurement outcomes ensures that in quantum mechanics 
the phrase `the outcome of a measurement of an observable $A$ at time $t'$
has an unambiguous meaning.
This may suggest the existence of a correlation between the measurement outcome, or the post-measurement value of the meter, and the pre-measurement value of the measured 
observable as a common cause for the coincidence of the outcomes.

Now we focus on the measurement carried out by the sole observer, to show that every probability reproducible
measurement of an observable is indeed ``value reproducible'' as precisely formulated below.

Let $A$ be the observable of the system $\bS$ to be measured.  
Let $M$ be the meter observable of the observer in the environment.
We assume that at time 0 the system $\bS$ is in an arbitrary state  
$\kps$ and the environment $\bE$ is in a fixed state $\kxi$.
In order to measure the observable $A$ at time 0, the observer locally measures the meter observable $M$
at time $\ta>0$.   
Then, the time evolution operator $U(\ta)$ of the total system $\bS+\bE$
determines the Heisenberg operators $A(0)$ and   $M(\ta)$, where
$A(0)=A\otimes I$,
$M(\ta)=U(\ta)^{\dagger}(I\otimes M) U(\ta)$.
We pose the following assumption.
The measurement satisfies the {\em probability reproducibility condition}, \ie,  
\deql{PRC}{ \Pr\{M(\ta)=x\}=\Pr\{A(0)=x\}}
for any vector state $\kps$ of $\bS$. 

Since quantum mechanics predicts a relation between values of observables only in the form 
of probability correlations, the coincidence between the pre-measurement value of the measured observable 
and the post-measurement value of the meter observable should be best expressed by the relation
  \deql{FM}{  
 \Pr\{A(0)=x,M(\ta)=y\}=0
 }
if $x\ne y$.
However, this relation shows a difficulty.
Since $A(0)$ and $M(\ta)$ may not
commute in general, the joint probability distribution may not  be well-defined.

In what follows, we shall show that under the probability reproducibility condition,
the above joint probability is actually well-defined to satisfy \Eq{FM}
as $A(0)$ and $M(\ta)$ commute on the subspace generated by the observables $A(0)$, $M(\ta)$ and the state $\kps$.

To see this, 
recall the notion of {\em partial commutativity} or 
{\em state-dependent commutativity} introduced 
by von Neumann \cite[p.~230]{vN32E}:
{\em If a state $\kPs$ is a superposition of 
common eigenstates $\ket{X=x,Y=y}$ of 
observables  $X$ and $Y$ of the form 
\deql{vN-1}{
\kPs=\sum_{(x,y)\in S} c_{x,y}\ket{X=x,Y=y},
}
where $S\subseteq \R^2$,
then the joint probability distribution $\Pr\{X=x,Y=y\}$ of $X$ and $Y$ in $\kPs$
is well-defined as}
\deql{vN-2}{
\Pr\{X=x,Y=y\}=
\left\{
\begin{array}{cl}
|c_{x,y}|^2 & \mbox{if $(x,y)\in S$},\\
0& \mbox{otherwise}.
\end{array}
\right.
}
In this case, $X$ and $Y$ actually commute on the subspace $\cM$ generated by
$\{\ket{X=x,Y=y}\}_{(x,y)\in S}$, and we say that {\em $X$ and $Y$ commute 
in the sate $\kPs$}.  
The observables $X$ and $Y$ can be simultaneously measured in the
state $\kPs$, and the joint probability distribution of the outcomes 
$X=x$ and $Y=y$ of the simultaneous measurements satisfies \Eq{vN-2}
\cite[p.~230--231]{vN32E}.
In this case,  $\kPs\in\cM$ and
\deq{
P^{X}(x)\And P^{Y}(y)\kPs=P^{X}(x)P^{Y}(y)\kPs=P^{Y}(y)P^{X}(x)\kPs,
}
where $\And$ denotes the infimum of two projections.
The joint probability distribution $\Pr\{X=x,Y=y\}$ satisfies
\deq{
\Pr\{X=x,Y=y\}=\av{\Ps|P^{X}(x)\And P^{Y}(y)|\Ps}
=\av{\Ps|P^{X}(x)P^{Y}(y)|\Ps}
=\av{\Ps|P^{Y}(y)P^{X}(x)|\Ps}
}
for any $x,y\in\R$. Moreover,
\deql{vN-3}{
\av{\Ps|f(X,Y)|\Ps}=\sum_{x,y}f(x,y)\Pr\{X=x,Y=y\}
}
for any real polynomial $f(X,Y)$ of $X$ and $Y$ \cite[Theorem 1]{19A1}.

Based on the above notion of the state-dependent commutativity due to von Neumann \cite{vN32E}, 
a precise formulation for the value reproducibility is given as follows.
A measurement of an observable $A$ described by the system-environment time evolution $U(\ta)$ from time $t=0$
to $t=\ta$ with the fixed initial environment state $\kxi$ is said to satisfy the {\em value reproducibility condition} if 
the pre-measurement observable $A(0)$ to be measured and the post-measurement meter observable $M(\ta)$ commute 
in the initial state $\kps\kxi$ and the joint probability distribution $\Pr\{A(0)=x,M(\ta)=y\}$ satisfies \Eq{FM} 
for any state vector $\kps$ of the measured system.
The terminology "pre-measurement value of the observable to be measured" and the "post-measurement
value of the meter observable" is defined through the well-defined joint probability distribution  $\Pr\{A(0)=x,M(\ta)=y\}$,
where $x$ is called the pre-measurement
value of the observable $A$ to be measured and $y$ is called the post-measurement value of the meter observable $M$.
Thus, the well-defined joint probability distribution in \Eq{FM} refers to the probability correlation
between the pre-measurement value of the observable to be measured and the post-measurement value of the meter observable.
 
 Then the following theorem holds.

\bTheorem[Value Reproducibility Theorem] \label{th:2}
Every probability reproducible measurement of an observable is value reproducible.
\eTheorem

A proof is given in the Methods section.

Now, we have seen that the proof of Theorem \ref{th:2} shows that (i) the time-like entanglement condition
implies the value reproducibility condition.
We already have seen that the proof of Theorem \ref{th:1} shows that (ii)  the probability reproducibility condition 
implies the time-like entanglement condition.
Thus, we have shown the following implication relations:
\deq{
\fbox{probability reproducibility condition} \Then \fbox{time-like entanglement condition} \Then \fbox{value reproducibility condition}  
}
We show that the above three conditions
are actually all equivalent. 

\bTheorem\label{th:E}
For any measurement of an observable $A$ in the state $\kps$ with the pre-measurement observable $A(0)$ and
the post-measurement meter observable $M(\ta)$ 
in the fixed environment state $\kxi$, the following conditions are all equivalent. 
\benum
\setlength{\leftskip}{18pt}
\item[(i)]
(Probability reproducibility condition) For any $x\in\R$ and any system state $\kps$,
\deql{PR}{
\av{\ps,\xi|P^{M(\ta)}(x)|\ps,\xi}=\av{\ps|P^{A}(x)|\ps}.
}
\item[(ii)]
(Time-like entanglement condition) For any $x\in\R$ and any system state $\kps$,
\deql{EFC}{
P^{M(\ta)}(x)\ket{\ps,\xi}=P^{A(0)}(x)\ket{\psi,\xi}.
}
\item[(iii)]
(Value reproducibility condition) For any system state  $\kps$,  the observables
$A(0)$ and  $M(\ta)$ commute in the initial state $\kps\kxi$ and satisfy
\deql{VR}{
\av{\ps,\xi|P^{A(0)}(x)P^{M(\ta)}(y)|\psi,\xi}=0
}
if $x\ne y$.
\eenum
\eTheorem
A proof is given in the Methods section.

Since the proof of Theorem \ref{th:1} have shown that the intersubjectivity condition is a straightforward consequence from the time-like entanglement condition for two space-like separated meter observables, we have eventually shown the following implication relations:
\beqa
\fbox{probability reproducibility condition} \Leftrightarrow &\hspace{-7pt}\fbox{time-like entanglement condition}\hspace{-7pt}&\Leftrightarrow \fbox{value reproducibility condition}  \nn\\
&\Downarrow&\\
&\fbox{intersubjectivity condition}&\nn
\eeqa

The implication `value reproducibility condition' $\Then$ `intersubjectivity
 condition' can also be obtained by the transitivity of perfect correlations 
\cite{06QPC}.  We say that two observables $X$ and $Y$ are perfectly correlated
in a state $\kPs$, and write $X=_{\kPs}Y$, if $\av{\Ps|P^{X}(x)P^{Y}(y)|\Ps}=0$ for $x\ne y$.
Then it is shown that perfect correlations are transitive, \ie,   $X=_{\kPs}Y$ and $Y=_{\kPs}Z$ implies  
$X=_{\kPs}Z$ for any observables $X,Y,Z$ \cite[Theorem 4.4]{06QPC}.
Now we suppose the value reproducibility conditions for the pairs $(A(0),M_1(\ta_1))$
and  $(A(0),M_2(\ta_2))$.  Then $M_1(\ta_1)=_{\ket{\ps}\ket{\xi}} A(0)$
and   $A(0)=_{\ket{\ps}\ket{\xi}} M_2(\ta_2)$, so that we obtain the intersubjectivity 
condition,  $M_1(\ta_1)=_{\ket{\ps}\ket{\xi}} M_2(\ta_2)$, by the transitivity of perfect correlations.

Under Assumptions 1 (locality) and 2 (probability reproducibility), 
we have found the perfect correlation between $A(0)$ and $M_1(\ta_1)$ and that between 
$A(0)$ and $M_2(\ta_2)$ from the Value Reproducibility Theorem. 
Then the perfect correlation between $M_1(\ta_1)$ and $M_2(\ta_2)$ found in the Intersubjectivity 
Theorem is considered to be a straightforward consequence of the transitivity of perfect correlations for the other two pairs.
Therefore we can conclude that the pre-measurement value of the observable to be measured
is a common cause for the perfect correlation between the outcomes of two or more space-like
separated observers, through the perfect correlations with the post-measurement meters
realized by the interaction with the environment.

\section*{Value reproducibility of the von Neumann model of repeatable measurements}
In the conventional approach to quantum measurements of an observable
$
A=\sum_a a\ket{\ph_a}\bra{\ph_a}
$
due to von Neumann \cite{vN32E},
the measurement is required to satisfy both the probability reproducibility condition, \Eq{PRC0}, 
and the repeatability condition, \Eq{RC},
whereas the rejection of the realism
in quantum mechanics has long been considered to defy 
the value reproducibility condition, \Eq{FM}, as  stated by Schr\"{o}dinger \cite{Sch35E}.
However,   we should point out that 
{\em the conventional analysis of measuring process given by \Eq{U-sp}
has failed to unveil the fact that the measurement actually satisfies
the value reproducibility condition}, as follows.

Now, we shall show that \Eq{U-sp} can also be rewritten as
\deql{t-ent}{
\ket{\psi}\ket{\xi}=\sum_{a}c_a \ket{A(0)=a,M(\ta)=a},
}
which implies that the joint probability distribution $\Pr\{A(0)=a,M(\ta)=m\}$ is well-defined and  
\deq{\Pr\{A(0)=a,M(\ta)=m\}=\de_{a,m}|c_a|^2,} 
so that the value reproducibility condition, \Eq{FM}, holds. 
To see this, note that we obtain the relations
\deq{
A(0)\ket{\ph_a}\ket{\xi}
&=(A\otimes I)\ket{\ph_a}\ket{\xi}
= a\ket{\ph_a}\ket{\xi},\\
M(\ta)\ket{\ph_a}\ket{\xi}
&= U(\ta)^{\da}(I\otimes M)U(\ta)\ket{\ph_a}\ket{\xi}
= U(\ta)^{\da}(I\otimes M)\ket{\ph_a}\ket{\xi_a}
= aU(\ta)^{\da}\ket{\ph_a}\ket{\xi_a}
= a\ket{\ph_a}\ket{\xi}
}
for all $a$.
Thus, we have 
\deq{
P^{M(\tau)}(a)P^{A(0)}(a)\ket{\ph_a}\ket{\xi}=P^{A(0)}(a)P^{M(\tau)}(a)\ket{\ph_a}\ket{\xi}=\ket{\ph_a}\ket{\xi}.
}
It follows that the state $\ket{\ph_a}\ket{\xi}$ is a joint eigenstate for observables $A(0)$ and $M(\tau)$
with the common eigenvalue $a$. 
Therefore,  \Eq{t-ent} follows with  $\ket{A(0)=a,M(\ta)=a}=\ket{\ph_a}\ket{\xi}$
 and we have proved that the value reproducibility condition, \Eq{FM}, holds.

 \section*{Intersubjectivity of some non-repeatable measurement models}
 
 It has long been claimed that a measurement entangles the measured observable with the meter, 
 or the measured system with the environment without any precise qualification.
However, it is wrong to consider that the intersubjectivity theorem is a consequence 
from this entangling nature of the measurement, 
since the entangling nature of the measurement is a consequence of the repeatability condition in addition 
to the probability 
reproducibility condition, whereas the intersubjectivity theorem is a consequence 
of the sole requirement of the probability reproducibility condition.
Thus, there is a measurement that does not entangle the measured system with the environment but 
satisfies the intersubjectivity condition, namely, entangles all the meters.

Recall that the unitary operator $U(\ta)$ for the von Neumann model given in \Eq{101103a}
has been shown to satisfy both repeatability condition and probability reproducibility condition.
To clarify the above point,  consider, now instead of $U(\ta)$, the unitary operator $V(\ta)$ satisifying
 \deql{240617y}{
V(\ta)\ket{\ph_a}\ket{\xi}=\ket{\ph}\ket{\xi_a},
}
where $\kph$ is an arbitrary but fixed state of the measured system.
Note that among the qubit measurements, the CNOT gate is a typical example of $U(\ta)$, 
while the SWAP gate is a typical example of $V(\ta)$. 

If the initial system state is a superposition
$\kps=\sum_{a}c_a \ket{\ph_{a}}$ then by linearity we obtain
\deql{V-sp}{
V(\tau)\ket{\psi}\ket{\xi}=\sum_{a}c_a\kph\ket{\xi_{a}}=\kph\Big(\sum_{a}c_a\ket{\xi_{a}}\Big).
}
Thus, this measurement makes no entanglement between the measured observable and the meter,
whereas this measurement satisfies the probability reproducibility condition by the relations
\deq{
\Pr\{M(\tau)=x\}=\| (I\otimes \ketbra{\xi_x})V(\tau)\ket{\psi}\ket{\xi}\|^2=
\Big|\Big(\ket{\xi_{x}},\sum_{a}c_a\ket{\xi_{a}}\Big)\Big|^2=|c_x|^2=\Pr\{A(0)=x\}.}

An interesting feature of the intersubjectivity theorem is that if such a unitary operator $V(\ta)$ is extended 
to two meters then it does not entangle the meters with the measured object either,  but entangles the two meters each other.  To see this, let $W(\ta)$ be such that
\deql{240617x}{
W(\ta)\ket{\ph_a}\ket{\xi}\ket{\xi'}=\kph\ket{\xi_a}\ket{\xi'_a},
}
which extends $V(\ta)$ to another meter $M'=\sum_{a}a \ketbra{\xi'_a}$ in the initial state $\ket{\xi'}$.
If the initial system state is a superposition
$\kps=\sum_{a}c_a \ket{\ph_{a}}$ then by linearity we obtain
\deql{W-sp}{
W(\tau)\ket{\psi}\ket{\xi}\ket{\xi'}=\sum_{a}c_a\kph\ket{\xi_{a}}\ket{\xi'_{a}}=\kph\Big(\sum_{a}c_a\ket{\xi_{a}}\ket{\xi'_a}\Big).}
Thus, this measurement does not entangle the measured observable with none of the two meters either, 
but both meters satisfy the probability reproducibility condition
\deq{
\Pr\{M(\ta)=x\}=\Pr\{M'(\ta)=x\}=|c_x|^2=\Pr\{A(0)=x\},}
and both are entangled each other to satisfy the intersubjectivity condition
\deq{
\Pr\{M(\ta)=x,M'(\ta)=y\}=0
}
if $x\ne y$.

Therefore, we conclude that the intersubjectivity theorem is not a consequence of the known type of entangling nature of measurements, while it reveals a new type of entangling nature of measurements that entangles all the 
meters and that is shared by all the probability reproducible measurements.

\section*{Unconventional generalized observables are not value reproducibly measurable}
In the preceding sections, we have shown that any probability reproducible measurement 
indeed reproduces the pre-measurement value, whether the repeatability is satisfied or not. 
It is an interesting problem to what extent a probability reproducible measurement
of a `generalized' observable can be value reproducible.  
Here,  we answer this question rather surprisingly: only the conventional observables can be measured 
value reproducibly. 

A generalized observable $A$ on a Hilbert space $\cH$ is 
called {\em value reproducibly measurable} 
if there exists a measuring process $(\cK,\kxi,U(\ta),M)$ 
for a Hilbert space $\cH$ satisfying condition (iii) of Theorem \ref{th:E}
(for the generalized observable $A$),
where $P^{A(0)}(x)=P^{A}(x)\otimes I$.
Then, we have 
\bTheorem\label{th:4}
A generalized observable is value reproducibly measurable 
if and only if it is an observable (in the conventional sense). 
\eTheorem
The proof is given in the Methods section.

\section*{Discussion}
Schr\"{o}dinger \cite[p.~329]{Sch35E} argued that a measurement does not ascertain
the pre-measurement value of the observable and is only required to be repeatable.
Since the inception of quantum mechanics, this view has long been supported
as one of the fundamental tenets of quantum mechanics. 
In contrast, we have shown in the Value Reproducibility Theorem (Theorem \ref{th:2}) that any probability reproducible measurement 
indeed reproduces the pre-measurement value, 
whether the repeatability is satisfied or not. 

It is an interesting problem to what extent a probability reproducible measurement
of a `generalized' observable can be value reproducible.  Theorem \ref{th:4}
answers this question rather surprisingly as that only conventional observables can be measured 
value-reproducibly.  This suggests a demand for more careful analysis on the notion 
of observables in foundations of quantum mechanics.
In this area, generalized probability theory \cite{80OG,Jan14} has recently been 
studied extensively.
However, the theory only has the notion of `generalized' observable, but
does not have the counter part of `conventional' observables being value-reproducibly 
measurable.

In this paper, we have considered the notion of measurement of observables `state-independently',
and we take it for granted that a measurement of an observable is accurate if and only
if it satisfies the probability reproducibility in all states.
However, this does not mean that `state-dependent' definition of an accurate measurement of
an observable should only require the probability reproducibility in a given state, since requiring the
probability reproducibility for all the state is logically and extensionally equivalent to requiring 
the value reproducibility for all the state as shown in this paper.
In the recent debate on the formulation of measurement uncertainty relations,
some authors have claimed that the state-dependent approach to this problem is not tenable,
based on the state-dependent probability reproducibility requirement \cite{BLW14,KJR14}.
In contrast, we have recently shown that state-dependent approach to 
measurement uncertainty relations is indeed tenable, based on the state-dependent 
value reproducibility requirement \cite{19A1}. 
The debate suggests that the value reproducibility is more reasonable requirement
for the state-dependent accuracy of measurements of observables. 

The cotextuality in assigning  the values to observables shown by the theorems 
due to Kochen-Specker \cite{KS67} and Bell \cite{Bel64} is often considered 
as the rejection of realism.  
However, it should be emphasized that what is real depends on a particular 
philosophical premise, and it is not completely determined by physics.
Here, we have revealed a new probability correlation, \Eq{Main}, predicted solely 
by quantum mechanics ensuring that the outcome of a measurement of an observable 
is unambiguously defined in quantum mechanics worth communicating intersubjectively.
Further, we have shown that the intersubjectivity of outcomes of measurements
is an immediate consequence from another new probability correlation, \Eq{FM},
the value-reproducibility of measurements. 
Since the value reproducibility is in an obvious conflict with the conventional understanding of 
`rejection of realism', 
it would be an interesting problem to interpret quantum reality taking into account
both the contextuality of value-assignments of observables and 
the intersubjectivity and the value reproducibility of outcomes of measurements.

In this connection, we have discussed the logical characterization of contextual hidden-variable 
theories based on the recent development of quantum set theory \cite{23A3}.
From this approach, we can conclude that the pre-measurement value of the observable 
to be measured is an element of reality in the context, in which the post-measurement 
meter observable is an element of reality to be actually read out by the observer,
as a consequence of the perfect correlation between them ensured by the value reproducibility theorem. 
A. Khrennikov \cite{Khr24} discussed  the interpretational implication of the intersubjectivity theorem to the QBism individual agent perspective.
We will further discuss interpretational issues of our results of the intersubjectivity and the value reproducibility of outcomes of quantum measurements elsewhere.

\section*{Methods}
\section*{Proof of Theorem \ref{th:1}}
\bProof
From Assumption 2, \Eq{A2}, we have
\deq{\|P^{M_1(\tau_1)}(x)\kps\kxi\|^2=\|P^{A(0)}(x)\kps\kxi\|^2.}
Since $\kps$ is arbitrary, replacing it
by $P^{A}(y)\kps/\|P^{A}(y)\kps\|$ if $P^{A}(y)\kps\ne 0$, we obtain
\deq{\|P^{M_1(\tau_1)}(x)P^{A(0)}(y)\kps\kxi\|^2
&=\|P^{M_1(\tau_1)}(x)(P^{A}(y)\kps)\kxi\|^2=\|P^{A(0)}(x)(P^{A}(y)\kps)\kxi\|^2\nn\\
&=\de_{x,y}\|P^{A(0)}(y)\kps\kxi\|^2.}
Since $P^{M_1(\tau_1)}(x)$ is a projection, it follows that 
\deq{P^{M_1(\tau_1)}(x)P^{A(0)}(y)\kps\kxi=\de_{x,y}P^{A(0)}(y)\kps\kxi.}
Summing up both sides of the above equation for all $y$, we obtain
\deql{QPC-1}{P^{M_1(\tau_1)}(x)\kps\kxi=P^{A(0)}(x)\kps\kxi.}
Similarly,
\deql{QPC-2}{P^{M_2(\tau_2)}(x)\kps\kxi=P^{A(0)}(x)\kps\kxi.}
Therefore, from Assumption 1, \Eq{JPD}, we have
\deql{QPC-3}{\Pr\{M_1(\tau_1)=x,M_2(\tau_2)=y\}
=\av{\psi,\xi|P^{M_1(\tau_1)}(x)P^{M_2(\tau_2)}(y)|\psi,\xi}=\av{\psi,\xi|P^{A(0)}(x)P^{A(0)}(y)|\psi,\xi}=0}
if $x\ne y$.
Thus, we conclude that the joint probability distribution of the outcomes of the simultaneous measurements of the observable $A$ by observers I and II satisfies the intersubjectivity condition, \Eq{Main},
which shows that the outcomes are always identical.
\eProof

\section*{Intersubjectivity for simultaneous position measurements}
{\bf Example 1.}
The system $\bS$ to be measured has canonically conjugate observables $Q,P$ 
on an infinite dimensional state space with
$[Q,P]=i\hbar$.  Consider the measurement of the observable $A=Q$.
The environment $\bE$ consists of  $n$ sets of canonically conjugate observables 
$Q_j,P_j$ with $[Q_j,P_k]=\de_{jk}$, $[Q_j,Q_k]=[P_j,P_k]=0$ for $j,k=1,\ldots,n$.  
Here, the part of the actual environment not effectively interacting with $\bS$ can be 
neglected without any loss of generality.
Consider $n$ observers with their meters $M_j=Q_j$ for $j=1,\ldots,n$.
Suppose that the system $\bS$ is in an arbitrary state $\kps$ and the environment $\bE$
is in the joint eigenstate $\kxi=\ket{Q_1=0,\ldots,Q_n=0}$.
The interaction between $\bS$ and $\bE$ is given by
\deq{
H=KQ\otimes (P_1+\cdots+P_n),
}
where the coupling constant $K$ is large enough to neglect the other term in the
total Hamiltonian of the composite system $\bS+\bE$.
Then, we have
\deqn{
\frac{d}{dt}Q_j(t)&= \frac{1}{i\hbar}[Q_j(t),H(t)]=KQ(t),\\
Q_j(t)&= Q_j(0)+KtQ(0).
}
Assumptions 1 and 2 are satisfied for $\tau_j=1/K$  with $j=1,\ldots,n$,
\ie,
\deqn{
[M_j(\tau_i),M_k(\tau_k)]&= 0,\\
\av{\psi,\xi|P^{M_j(\tau_j)}(x)|\psi,\xi}dx&= \av{\psi|P^{Q}(x)|\psi}dx.
}
In this case, $M_j(t)-M_k(t)=Q_j(t)-Q_k(t)$ 
are the constant of the motion for all $j,k$, \ie, 
\deqn{
\frac{d}{dt}(Q_j(t)-Q_k(t))&= \frac{1}{i\hbar}[Q_j(t)-Q_k(t),H(t)]
=0.
}
Thus, the outcomes are identical for all the observers, \ie, 
\deq{
\av{\psi,\xi|P^{M_j(\tau_j)}(x)P^{M_k(\tau_k)}(y)|\psi,\xi}dx\,dy
&= \de(x-y)|\psi(x)|^2dx\,dy.
}

\section*{Non-Intersubjectivity of unconventional generalized observables}

{\bf Example 2.}
Let $A$ be a generalized observable on a system $\bS$ 
described by a Hilbert space $\cH$
such that $P^{A}(x)=\mu(x)I$,
where $\mu$ is a probability distribution, \ie, $\mu(x)\ge 0$ for all $x\in\R$ and $\sum_{x\in\R}\mu(x)=1$.
Let $X=\{x\in\R\mid\mu(x)>0\}$.
Suppose that the environment $\bE$ consists of  two subsystems so that
the environment is described by a Hilbert space $\cK=\cL\otimes\cL$,
where $\cL$ is a Hilbert space spanned by an orthonormal basis 
$\{\ket{x}\}_{x\in X}$.
Suppose that observers I and II 
measure the meter observables $M_1(\ta_1)=I\otimes M\otimes I$ and $M_2(\ta_2)=I\otimes I\otimes M$  on $\cH\otimes\cK$, respectively, 
which may be constants of motion,
where $M=\sum_{x}x \ketbra{x}$,
so that Assumption 1 is satisfied.
The initial state of the  environment $\bE$ can be represented by
\deq{
\kxi=\sum_{x,y\in X}c_{x,y}\ket{x}\ket{y}.
}
Then, the joint probability distribution of $M_1(\ta_1)$ and $M_2(\ta_2)$ is
given by
\deqn{
\Pr\{M_1(\ta_1)=x,M_2(\ta_2)=y\}
= \av{\ps,\xi|P^{M_1(\ta_1)}(x)P^{M_2(\ta_2)}(y)|\ps,\xi}=|c_{x,y}|^2.
}
Thus, Assumption 2 is satisfied if and only if 
$\mu(x)=\sum_{y}|c_{x,y}|^2=\sum_{y}|c_{y,x}|^2$.
Thus, under Assumptions 1 and 2, the joint probability distribution
$\mu(x,y)=\Pr\{M_1(\ta)=x,M_2(\ta)=y\}$ can be an arbitrary 
2-dimensional probability distribution such that 
$\sum_{y}\mu(x,y)=\sum_{y}\mu(y,x)=\mu(x)$.
In this case,  \Eq{Main} is satisfied if and only if  
$|c_{x,y}|^2=\de_{x,y}\mu(x)$.
Thus, we have continuously parametrized models with $\mu(x)$ and $c_{x,y}$ for which 
Assumptions 1 and 2 are satisfied, \ie,  $\mu(x)=\sum_{y}|c_{x,y}|^2=\sum_{y}|c_{y,x}|^2$  but the intersubjectivity condition,  \Eq{Main},  is not satisfied, \ie, 
$|c_{x,y}|^2\ne\de_{x,y}\mu(x)$ for some $(x,y)\in X^{2}$.

\section*{Proof of Theorem \ref{th:2}}

\bProof
Under the probability reproducibility condition, we can adapt the proof of Theorem 1 for $M(\ta)=M_1(\ta_1)$
to obtain the time-like entanglement condition 
\deql{QPC-1A}{
P^{M(\ta)}(x)\kps\kxi=P^{A(0)}(x)\kps\kxi.
}
Let $S=\{(x,y)\in\R^2\mid P^{A(0)}(x)P^{M(\ta)}(y)\kps\kxi\ne 0\}$, and 
$c_{x,y}=\|P^{A(0)}(x)P^{M(\ta)}(y)\kps\kxi\|$.
For any $(x,y)\in S$, define $\ket{A(0)=x,M(\ta)=y}$ by
\deq{\ket{A(0)=x,M(\ta)=y}&={c_{x,y}^{-1}}P^{A(0)}(x)P^{M(\ta)}(y)\kps\kxi.
}
From \Eq{QPC-1A},
\deql{JES}{P^{A(0)}(x)P^{M(\ta)}(y)\kps\kxi&=P^{A(0)}(x)P^{A(0)}(y)\kps\kxi=\de_{x,y}P^{A(0)}(x)\kps\kxi
=\de_{x,y}P^{M(\ta)}(x\kps\kxi)\nn\\
&=P^{M(\ta)}(y)P^{M(\ta)}(x)\kps\kxi=P^{M(\ta)}(y)P^{A(0)}(x)\kps\kxi.}
Thus, $\ket{A(0)=x,M(\ta)=y}$ is a common eigenstate for $A(0)$ with eigenvalue $x$ and $M(\ta)$ with eigenvalue $y$
 such that
\deq{\kps\kxi=\sum_{(x,y)\in\R^{2}}P^{A(0)}(x)P^{M(\ta)}(y)\kps\kxi=
\sum_{(x,y)\in S}c_{x,y}\ket{A(0)=x,M(\ta)=y}.} 
Thus, the joint probability distribution $\Pr\{A(0)=x,M(\ta)=y\}$
is well defined as 
\deq{\Pr\{A(0)=x,M(\ta)=y\}=\|P^{A(0)}P^{M(\ta)}(x)\kps\kxi\|^2.}
From the second equality in \Eq{JES}, we have the value reproducibility condition, $\Pr\{A(0)=x,M(\ta)=y\}=0$ if $x\ne y$.
\eProof

\section*{Proof of Theorem \ref{th:E}}
\bProof
The proofs for the implications (i)$\Then$(ii) and (ii)$\Then$(iii) have been given in the proofs of 
Theorems \ref{th:1} and \ref{th:2} as discussed after the proof of Theorem \ref{th:2}, and hence
it suffices to prove the implication  (iii)$\Then$(i).
From (iii) we have
\deq{
& \av{\ps|P^{A}(x)|\ps}=\Pr\{A(0)=x\}=\sum_{y}\,\Pr\{A(0)=x,M(\ta)=y\}=|c_{x,x}|^2,\\
& \av{\ps,\xi|P^{M(\ta)}(x)|\ps,\xi}=\Pr\{M(\ta)=x\}=\sum_{y}\,\Pr\{A(0)=y,M(\ta)=x\}=|c_{x,x}|^2,
}
and hence the probability reproducibility condition (i) follows.
\eProof

\section*{Proof of Theorem \ref{th:4}}
\bProof
In the Result section it has been shown that every observable is value reproducibly 
measurable. It suffices to show the converse.
Let $\kps\in\cH$ be a state vector.
Let $A$ be a generalized observable on $\cH$, which is value reproducibly measurable
with a measuring process $(\cK,\kxi,U(\ta),M)$ satisfying \Eq{VR} if $x\ne y$.
By the Naimark-Holevo dilation theorem \cite{Hol01}, there exist a Hilbert space $\cG$, 
a state vector $\ket{\eta}\in\cG$, and an observable $B$ on $\cG\otimes \cH$ 
such that
\deq{
P^{A}(x)=\av{\eta|P^{B}(x)|\eta} 
}
for all $x$.  Let $B(0)=B\otimes I_{\cK}$ and
$N(\ta)=I_{\cG}\otimes M(\ta)$. 
Let $\ket{\eta_1},\ket{\eta_2},\ldots$ be an orthonormal basis of $\cG$
such that $\ket{\eta}=\ket{\eta_1}$.
Let $\ket{\psi_1},\ket{\psi_2},\ldots$ be an orthonormal basis of $\cH$
such that $\kps=\ket{\psi_1}$.
Let $\ket{\xi_1},\ket{\xi_2},\ldots$ be an orthonormal basis of $\cK$
such that $\kxi=\ket{\xi_1}$.
 Then, we have
\deq{
\av{\eta_j,\psi_k,\xi_l|P^{N(\ta)}(y)|\eta,\psi,\xi}=0
}
if $j\not=1$.  Thus, we have
\deql{joint-dilation}{
\av{\eta,\psi,\xi|P^{B(0)}(x)P^{N(\ta)}(y)|\eta,\psi,\xi}
&=
\sum_{j,k,l}
\av{\eta,\psi,\xi|P^{B(0)}(x)|\eta_j,\psi_k,\xi_l}\av{\eta_j,\psi_k,\xi_l|P^{N(\ta)}(y)|\eta,\psi,\xi}\nn\\
&=
\sum_{k,l}\av{\eta,\psi,\xi|P^{B(0)}(x)|\eta,\psi_k,\xi_l}\av{\eta,\psi_k,\xi_l|P^{N(\ta)}(y)|\eta,\psi,\xi}\nn\\
&=
\sum_{k,l}\av{\psi,\xi|P^{A(0)}(x)|\psi_k,\xi_l}\av{\psi_k,\xi_l|P^{M(\ta)}(y)|\psi,\xi}\nn\\
&=
\av{\psi,\xi|P^{A(0)}(x)P^{M(\ta)}(y)|\psi,\xi}.
}
Let $\kPs=\ket{\eta}\kps\kxi$.
From \Eq{joint-dilation} and the value reproducibility of $A$, we have
\deql{PC-BN}{
\av{\Ps|P^{B(0)}(x)P^{N(\ta)}(y)|\Ps}=0.
}
if $x\ne y$.
It follows that 
\begin{align}
\av{\Ps|P^{B(0)}(x)P^{N(\ta)}(x)|\Ps}
=\|P^{B(0)}(x)\kPs\|^2
=\|P^{N(\ta)}(x)\kPs\|^2,
\end{align}
and hence
\deq{
\|P^{B(0)}(x)\kPs-P^{N(\ta)}(x)\kPs\|^2
=
\|P^{B(0)}(x)\kPs\|^2+\|P^{N(\ta)}(x)\kPs\|^2
-2{\rm Re}\av{\Ps|P^{B(0)}(x)P^{N(\ta)}(x)|\Ps}
=0
}
Therefore, we obtain
\deq{
P^{B(0)}(x)\ket{\eta}\kps\kxi=P^{N(\ta)}(x)\ket{\eta}\kps\kxi,
}
so that
\deql{s-ent-1}{
P^{A(0)}(x)\kps\kxi
&=\av{\eta|P^{B(0)}(x)|\eta}\kps\kxi
=\av{\eta|P^{N(\ta)}(x)|\eta}\kps\kxi
=P^{M(\ta)}(x)\kps\kxi.
}
Since $\kps$ is arbitrary, replacing $\kps$ in \Eq{s-ent-1} by
$P^{A}(x)\kps/\|P^{A}(x)\kps\|$ if $P^{A}(x)\kps\ne 0$, we have
\deql{s-ent-2}{
P^{A(0)}(x)(P^{A}(x)\kps)\kxi
=P^{M(\ta)}(x)(P^{A}(x)\kps)\kxi,
}
which holds even if $P^{A}(x)\kps= 0$.
From \Eq{s-ent-1} and \Eq{s-ent-2}, we have
\deqn{
P^{A(0)}(x)^2\kps\kxi 
&=  P^{A(0)}(x)(P^{A}(x)\kps)\kxi
=   P^{M(\ta)}(x)(P^{A}(x)\kps)\kxi
=   P^{M(\ta)}(x)P^{A(0)}(x)\kps\kxi\nn\\
&=  P^{M(\ta)}(x)^2\kps\kxi
= P^{M(\ta)}(x)\kps\kxi
= P^{A(0)}(x)\kps\kxi.
}
Therefore, we have 
\deq{
P^{A}(x)^2\kps=P^{A}(x)\kps
}
for all state vector $\kps\in\cH$, so that $A$ is an observable. 
\eProof

\section*{\Large\bf Acknowledgements}
The author thanks Andrei Khrennikov for valuable discussions on an earlier version of this paper.
 The author acknowledges the supports of JSPS KAKENHI Grant Numbers JP24H01566, JP22K03424, and JST CREST Grant Number JPMJCR23P4, Japan.
\section*{\rm Author contributions}
The author researched, collated, and wrote this paper.
\section*{\rm Data availability}
No datasets were generated or analyzed during the current study.
\section*{\bf\em Competing interests}
The author has no competing interests relevant to this work.

\end{document}